\documentclass{article}
\usepackage[utf8]{inputenc}
\usepackage{geometry}
\usepackage{xcolor}
\usepackage{hyperref}
\usepackage{amsmath}
\usepackage{amssymb}
\usepackage{amsfonts}
\usepackage{enumerate}
\usepackage{comment}
\usepackage{xpatch}
\usepackage{graphicx}
\usepackage{tabularx}
\usepackage{braket}
\usepackage{physics}
\usepackage{amsthm}
\usepackage{tikz}
\usepackage{commath}
\usepackage{mathtools}
\usepackage{qcircuit}
\usepackage{algorithm}
\usepackage{algorithmicx}
\usepackage{algpseudocode}
\usepackage{tabto}
\usepackage{pgf-umlsd}
\usepackage{bm}
\usepackage{multirow, multicol}
\usepackage{float}
\usepackage{pdfpages}
\usepackage{caption}
\usepackage{subcaption}
\usepackage{authblk}
\usepackage{cite}

\newcommand{\poly}{\mathrm{poly}}
\hypersetup{
    colorlinks=true,
    linkcolor=blue,
    filecolor=magenta,      
    urlcolor=cyan,
    pdfpagemode=FullScreen,
    citecolor=blue
    }
    
\newcounter{protocol}
\makeatletter

\newtheorem{theorem}{Theorem}
\newtheorem{lemma}{Lemma}

\theoremstyle{remark}

\theoremstyle{definition}
\newtheorem{definition}{Definition}

\newcommand{\phidep}{\Phi_{\mathrm{dep}}}

\usepackage{subfiles}
\title{\Large Noise-tolerant learnability of shallow quantum circuits from statistics and the cost of quantum pseudorandomness
}

\makeatletter
\def\thanks#1{\protected@xdef\@thanks{\@thanks
        \protect\footnotetext{#1}}}
\makeatother

\author[1,*]{Chirag Wadhwa\thanks{$^*$chirag.wadhwa@ed.ac.uk}}
\author[1]{Mina Doosti}

\affil[1]{\small \emph{School of Informatics, University of Edinburgh, United Kingdom}}
\date{}

\begin{document}

\maketitle

\begin{abstract}
    In this work, we study the learnability of quantum circuits in the near term. We demonstrate the natural robustness of quantum statistical queries for learning quantum processes, motivating their use as a theoretical tool for near-term learning problems. We adapt a learning algorithm for constant-depth quantum circuits to the quantum statistical query setting, and show that such circuits can be learned in our setting with only a linear overhead in the query complexity. We prove average-case quantum statistical query lower bounds for learning, within diamond distance, random quantum circuits with depth at least logarithmic and at most linear in the system size. Finally, we prove that pseudorandom unitaries (PRUs) cannot be constructed using circuits of constant depth by constructing an efficient distinguisher using existing learning algorithms. To show the correctness of our distinguisher, we prove a new variation of the quantum no free lunch theorem.
\end{abstract}
\section{Introduction}
\label{sec:intro}
The problem of learning quantum processes is a fundamental question in quantum information, learning theory, benchmarking and cryptography among several other areas. However, tomography of quantum processes in general is known to require an exponential amount of data and computational time~\cite{bisio2009optimal,haah2023query}. Even so, many advances have been made to improve the efficiency of learning processes by focusing either on specific classes of quantum processes with special structure~\cite{lai2022learning,zhao2023learning,huang2024learning,nadimpalli2024pauli,low2009learning,chen2023testing,bao2023testing,torlai2023quantum,montanaro2008quantum,rouze2024quantum,harper2020efficient,harper2021fast,flammia2020efficient,chen2022quantum,flammia2021pauli,fawzi2023lower,chen2024tight} or by focusing on learning specific properties of processes rather than fully characterizing them~\cite{kunjummen2023shadow,huang2023learning, levy2024classical, chen2024predicting,chung2018sample}. An important feature to consider is the depth of the quantum circuit, especially in the near term where devices are prone to noise and cannot be used to reliably implement deep circuits. The utility of shallow quantum circuits has been widely studied~\cite{bravyi2018quantum,bravyi2020quantum}, making the learnability of shallow quantum circuits an interesting area of research, which has already been explored in~\cite{zhao2023learning,huang2024learning, nadimpalli2024pauli}.

Another important factor to consider in the near term is the prevalence of noise. Characterizing and correcting the noise of quantum devices are active areas of research that are crucial for the progress of the field of quantum computing~\cite{eisert2020quantum,harper2020efficient,cai2023quantum}. In classical learning theory, the study of learnability under noise was initiated by the work of Kearns~\cite{kearns1998efficient}, who proposed the statistical query model as a naturally robust model, as an algorithm that can learn a function using only statistical properties can be implemented robustly in a noisy setting. Statistical queries have also been explored in quantum learning theory, with recent work studying quantum statistical queries for learning many properties of quantum states and processes~\cite{arunachalam2021quantum,arunachalam2024role,nietner2023average,nietner2023unifying,angrisani2023learning,wadhwa2024learning}. In this work, we will focus on quantum statistical queries for learning quantum processes, and show their applicability in the development of robust algorithms. More specifically, our focus will be on the \emph{learnability of shallow quantum circuits from quantum statistical queries}, for which we will prove both upper and lower bounds. Our work signifies the use of quantum statistical queries as a powerful theoretical tool for studying the learnability of quantum processes in the near term.

Another field closely tied to learning theory is cryptography. Even though the techniques used in both fields are quite different, results in cryptography are often proven using tools from learning theory~\cite{pietrzak2012cryptography,alani2019applications} and vice-versa~\cite{zhao2023learning,jerbi2023shadows}. An area of research in quantum cryptography that has attracted much attention over the past few years is the notion of quantum pseudorandomness~\cite{ji2018pseudorandom}, where one hopes to replace truly random quantum objects with deterministic objects that are computationally indistinguishable from truly random ones. A line of recent work~\cite{metger2024simple,chen2024efficient} has focused on proving a secure construction for pseudorandom unitaries. While no fully secure construction has been found so far, these constructions do satisfy weaker notions of security. On the other hand, there has also been recent work showing necessary properties that any construction for pseudorandom unitaries must satisfy~\cite{haug2023pseudorandom,grewal2022low}. Studying such properties and analyzing the required resources is an important direction to develop efficient and secure constructions for PRUs. Towards this goal, we will use results from learning theory to prove that \emph{pseudorandom unitaries cannot be implemented by shallow circuits.} 
\subsection{Our contributions}
\paragraph{Quantum Statistical Query oracles:} Quantum statistical query (QSQ) access models have previously been studied for the tasks of learning quantum states~\cite{arunachalam2021quantum,arunachalam2024role,nietner2023unifying} and quantum processes~\cite{nietner2023average,angrisani2023learning,wadhwa2024learning}. Recently, a new multi-copy QSQ oracle was introduced in~\cite{nietner2023unifying}. Expanding on these definitions, we propose two new quantum statistical query oracles. First, we define a \textit{multi-copy quantum statistical query oracle for the task of learning a quantum process}. With this oracle, we allow the learner to query a $k$-register state, to which $k$-copies of the process are applied in parallel, followed by a simultaneous measurement over the $k$ registers. The learner is allowed to query states that are entangled across the $k$ registers and observables entangled across the registers as well. This oracle allows us to model learners that have stronger capabilities compared to the single-copy setting for processes but are still weaker than learners that can make general black-box queries. Next, we introduce a \textit{QSQ oracle for learning an observable}. Here, one queries the unknown observable with a state and receives an approximation of the expectation value. This oracle can be viewed as modelling the behaviour of an unknown quantum apparatus that takes copies of a state, performs some measurements and returns a classical expectation value. In this context, the task of learning the observable is equivalent to identifying the physical quantity the apparatus measures and can be used for benchmarking its behaviour. We define these oracles in Section~\ref{sec:oracles}.
\paragraph{Noise-tolerance of QPSQ learners:} The main motivation behind statistical query learning in the classical learning theory setting has been to provide a framework to study learnability robust to classification noise. The need for developing robust learners is even more crucial in the quantum setting, particularly in the near-term. In this direction, the robustness of QSQ learners for quantum states to various kinds of noise was shown in~\cite{gollakota2022hardness}. We extend this argument to quantum statistical queries for quantum processes (QPSQs), by showing that learners with access to this oracle are naturally robust to noise within an acceptable threshold. Further, we show that for global depolarizing noise, there is an efficient method to estimate the deviation of the noisy channel from the original one. This method only uses a single 2-copy QPSQ query. Moreover, we argue that using a 2-copy query is necessary, as the closely related problem of purity estimation is known to require exponentially many single-copy queries \cite{arunachalam2024role}. We present these results in Section~\ref{sec:robust}.
\paragraph{QPSQ learner for shallow quantum circuits:} We analyze the learning algorithm for constant-depth quantum circuits proposed in~\cite{huang2024learning}, and show that this algorithm can be adapted, with a linear overhead in the system size, to the $\mathsf{QPStat}$ access model. Along with our framework for developing robust algorithms, this implies a provably robust method to learn constant-depth quantum circuits. We present this result in Section~\ref{sec:learner}.
\paragraph{Lower bounds for learning logarithmic-to-linear depth random quantum circuits:}  Here, we show an \textit{average-case} query-complexity lower bound for learning brickwork random quantum circuits with depth logarithmic to linear in the system size. In this depth regime, our lower bound scales exponentially with the depth, providing a smooth characterization of the learnability of random shallow quantum circuits with respect to their depth. In comparison to the lower bound in~\cite{huang2024learning}, which showed an exponential (in system size) lower bound in the \emph{worst-case} for learning log-depth circuits, our result does not rule out efficient learnability of such circuits \textit{on average}. We prove this lower bound in Section~\ref{sec:hardness}.
\paragraph{Cost of quantum pseudorandomness:} Alongside a learning algorithm for constant-depth circuits, Huang \textit{et al}.~\cite{huang2024learning} showed an algorithm for verifying the output of their learning algorithm within an average distance $d_{\mathrm{avg}}$. We use both the learning and verification algorithms from~\cite{huang2024learning} in a black-box manner to construct an efficient distinguisher between Haar random unitaries and unitaries implementable by constant-depth circuits. As a result, we prove that \textit{constant-depth circuits cannot form pseudorandom unitaries (PRUs)}. Along the way, we extend the quantum no-free lunch theorem from~\cite{poland2020no}, and show an average-case hardness result for learning Haar-random unitaries with black-box access within bounded $d_{\mathrm{avg}}$. We present these results in Section~\ref{sec:pru}.
\subsection{Related work}

\paragraph{Quantum Statistical Query Learning:} Quantum statistical queries for learning quantum processes were proposed by~\cite{wadhwa2024learning}, where the authors showed average-case query complexity lower bounds for learning unitaries within bounded diamond distance. Nadimpalli \textit{et al.}~\cite{nadimpalli2024pauli} also consider a similar access model and show a low-degree learning algorithm in this model, allowing them to learn $\mathsf{QAC}^0$ channels with limited auxiliary qubits. The lower bounds of \cite{wadhwa2024learning} rely on a reduction from a many-versus-one distinguishing task, a technique introduced for classical statistical query learning by~\cite{feldman2017general}, and further used in QSQ learning by~\cite{nietner2023average,arunachalam2024role,nietner2023unifying}. Multi-copy QSQs for learning quantum states were also introduced in \cite{nietner2023unifying}. \cite{nietner2023average} also proved average-case lower bounds for learning \emph{output distributions} of shallow quantum circuits, under a weaker access model than the one we consider in this work. The lower bound in this work and that of~\cite{nietner2023average} for shallow quantum circuits use techniques developed in~\cite{hunter2019unitary,barak2020spoofing} to bound second-order moments of random shallow circuits.

\paragraph{Learning Unknown Observables:} The task of learning unknown observables was recently considered by Molteni \textit{et al}. in~\cite{molteni2024exponential}. We note a distinction between the access model considered in~\cite{molteni2024exponential} and our proposed oracle $\mathsf{QStat}_O$ (see Definition~\ref{def:qstato}). Specifically, Molteni \textit{et al}. consider learners with access to random examples of the form $(|\psi_x\rangle, \alpha)$, where $|\psi_x\rangle$ is a classically-described quantum state w.r.t a classical string $x$, and $\alpha$ is an estimation of the expectation value of the unknown observable on this state. On the other hand, $\mathsf{QStat}_O$ allows the learner to query any state $\rho$ of their choice, possibly adaptively, and returns an estimate of $\Tr(O\rho)$. Thus, the access model we consider here is stronger than the one in~\cite{molteni2024exponential}.

\paragraph{Quantum No Free Lunch Theorems:} We have shown an average-case lower bound for learning Haar-random unitaries. The original quantum no-free-lunch theorem (QNFLT) from~\cite{poland2020no} gave a similar result, where they lower bounded the error of any learning algorithm on average over both the Haar measure as well as a \emph{randomly sampled dataset}, and when the hypothesis is a unitary. On the other hand, our result holds even when the learner can make \emph{adaptive queries}. Other QNFLT variations include a worst-case lower bound~\cite{zhao2023learning} and bounds when one allows input states to be entangled with ancillary registers or mixed~\cite{zhao2023learning, sharma2022reformulation}.

\paragraph{Hardness of Learning and Pseudorandom Unitaries:} In \cite{yang2023complexity}, the authors show average-case hardness for learning Haar-random unitaries as well as pseudorandom unitaries, implying that efficiently learnable classes of unitaries cannot be pseudorandom. However, these results only hold against algorithms whose hypotheses are unitary channels (or close to one). Consequently, these results do not immediately imply that constant-depth quantum circuits cannot be pseudorandom, as the hypothesis of the learning algorithm of~\cite{huang2024learning} need not be a unitary channel. While our average-case lower bound for learning Haar-random unitaries is weaker than the one showed by~\cite{yang2023complexity}, it \emph{also holds against algorithms whose hypotheses are arbitary CPTP maps}. This is a key component of our no-go result for constant-depth PRUs.
 
 \paragraph{Note Added:} Since the first version of this work, a new result by~\cite{schuster2024random} showed that PRUs can be constructed optimally in depth $\mathsf{poly}(\log \log n)$, superseding our result in Theorem~\ref{thm:pru}.
\subsection{Structure of the paper}
In Section~\ref{sec:prelim}, we will introduce some preliminary material. In Section~\ref{sec:oracles}, we recall previously defined QSQ oracles, and introduce the two new oracles defined in this work. In Section~\ref{sec:robust}, we demonstrate the natural robustness of QPSQ algorithms as well as a method to benchmark depolarizing noise. In Section~\ref{sec:learner}, we will show a QPSQ learning algorithm for constant-depth circuits. In Section~\ref{sec:hardness}, we prove a depth-dependent lower bound for learning random quantum circuits in logarithmic to linear depth regimes. In Section~\ref{sec:pru}, we will show that constant-depth circuits cannot be used to construct pseudorandom unitaries.

\section{Preliminaries}
\label{sec:prelim}
For basic definitions of quantum computation and information, we refer the reader to~\cite{nielsen2010quantum}. We denote the $N \times N$ identity matrix as $I_N$ and we may omit the index $N$ when the dimension is clear from the context. We will write $\mathcal{M}_{N,N}$ to denote the set of linear operators from $\mathbb{C}^N$ to $\mathbb{C}^N$ and we define the set of quantum states as $\mathcal{S}_N \coloneqq \left\{\rho \in \mathcal{M}_{N,N} : \rho \succeq 0, \Tr[\rho] = 1\right\}$. We denote by $U(N)$ the group of $N$-dimensional unitary operators. We denote by $H_{N,N}$ the set of $N \times N$ Hermitian operators. For a unitary operator $U$, we may denote the corresponding channel $\mathcal{U} \coloneqq U(.)U^{\dag}$ without explicit definition. We now include some important definitions in quantum information that will be useful throughout.
\begin{definition}[Pauli operators]
    \label{def:paulis}
    The set of Pauli operators is given by
    \begin{equation}
        X = \begin{pmatrix} 0 & 1 \\ 1 & 0\end{pmatrix}, Y = \begin{pmatrix} 0 & -i \\ i & 0\end{pmatrix}, Z = \begin{pmatrix} 1 & 0 \\ 0 & -1\end{pmatrix}.     
    \end{equation}
    The set $\mathcal{P}_1=\{I,X,Y,Z\}$ forms an orthonormal basis for $\mathcal{M}_{2,2}$ with respect to the Hilbert-Schmidt inner product. We will refer to the set of tensor products of Pauli operators and the identity, i.e. the operators of the form $P\in \{I,X,Y,Z\}^{\otimes n}:=\mathcal{P}_n$ as \textit{Pauli strings} over $n$ qubits.
\end{definition}
\begin{definition}[Single-qubit Pauli eigenstates]
    \label{def:stab1}
    We define the set of eigenstates of the single-qubit Pauli operators as
    \begin{equation}
        \mathrm{stab}_1 = \{ |0\rangle, |1\rangle, |+\rangle, |-\rangle, |+y\rangle, |-y\rangle \},
    \end{equation}
    where $|0\rangle$ \& $|1\rangle$ are the eigenstates of $Z$ , $|+\rangle$ \& $|-\rangle$ are the eigenstates of $X$ and $|+y\rangle$ \& $|-y\rangle$ are the eigenstates of $Y$.
\end{definition}
\begin{definition}[Quantum Channels]
    A map $\mathcal{E}: \mathcal{S}_N \rightarrow \mathcal{S}_N$ is said to be completely positive if for any positive operator $A \in \mathcal{M}_{N^2,N^2} , (\mathcal{E} \otimes I)(A)$ is also a positive operator. $\mathcal{E}$ is said to be trace-preserving if for any input density operator $\rho, \Tr(\mathcal{E}(\rho)) = \Tr(\rho) = 1$. A quantum process $\mathcal{E}$ is defined as a Completely Positive Trace-Preserving (CPTP) map from one quantum state to another. We may use the terms quantum process and quantum channel interchangeably.
\end{definition}
\begin{definition}[Maximally Depolarizing Channel]
    The maximally depolarizing channel $\phidep$ acting on states in $S_N$ is defined as follows:
    \begin{equation}
        \phidep(\rho) = \Tr(\rho) \cdot \frac{I}{N}.
    \end{equation}
\end{definition}
\subsection{Quantum Distances}
Now, we define distances and accuracy measures for quantum states.
\begin{definition}[Trace Distance]
    The trace distance between two quantum states is given by
    \begin{equation}
        d_{\mathrm{tr}}(\rho, \sigma) = \frac{1}{2}\norm{\rho -\sigma}_1,
    \end{equation}
    where $\|\cdot\|_1$ is the Schatten 1-norm.
\end{definition}
\begin{definition}[{Fidelity}]
    The fidelity between two quantum states $\rho$ and $\sigma$ is given by 
    \begin{equation}
        F(\rho,\sigma) = \Tr\left(\sqrt{\sqrt{\rho} \sigma \sqrt{\rho} }\right)^2.
    \end{equation}
    In particular, when both states are pure, the fidelity can be written as
    \begin{equation}
        F(|\psi\rangle\langle\psi|, |\phi\rangle\langle\phi|) = \left|\langle \psi|\phi \rangle\right|^2. 
    \end{equation}
    Further, when at least one of the states is pure,
    \begin{equation}
        F(|\psi\rangle\langle\psi|, \rho) = \langle\psi|\rho|\psi\rangle,
    \end{equation}
    and we also have the following relation between the fidelity and trace distance.
    \begin{equation}
        1-F(|\psi\rangle\langle\psi|, \rho) \leq d_{\mathrm{tr}}(|\psi\rangle\langle\psi|, \rho).
    \end{equation}
\end{definition}
Next, we define the two distances between quantum channels that we will consider in this work. The first one is the diamond distance, which is a worst-case distance over all states, while the other distance is the average infidelity of the output states of the channels over Haar-random inputs.
\begin{definition}[Diamond norm and diamond distance]
    \label{def:diamond}
    For a quantum process $\mathcal{E}: \mathcal{S}_N \rightarrow \mathcal{S}_N$, and $\mathcal{I}$ the identity superoperator acting on $\mathcal{M}_{N,N}$, we define the diamond norm $\|\cdot\|_\diamond$
    \begin{equation}
        \norm{\mathcal{E}}_\diamond = \underset{\rho \in \mathcal{S}_{N^2}}{\mathrm{max}} \norm{(\mathcal{E} \otimes \mathcal{I})(\rho)}_1.
    \end{equation}
    We then define the diamond distance, $d_\diamond$, as
    \begin{equation}
        d_\diamond(\mathcal{E}_1, \mathcal{E}_2) = \frac{1}{2}\norm{\mathcal{E}_1 - \mathcal{E}_2}_\diamond.
    \end{equation}
\end{definition}
\begin{definition}[Average distance]
    \label{def:avg-dist}
    We define the average distance between two channels $d_{\mathrm{avg}}$ as the infidelity of the output states on average over Haar-random inputs (see Definition~\ref{def:haar}).
    \begin{equation}
        d_{\mathrm{avg}}(\mathcal{E}_1,\mathcal{E}_2) = 1 - \underset{|\psi\rangle \sim \mu_S}{\mathbf{E}}\left[F\left(\mathcal{E}_1(|\psi\rangle\langle\psi|), \mathcal{E}_2(|\psi\rangle\langle\psi|) \right)\right] .
    \end{equation}
\end{definition}
Next, we state a useful relation between the average and diamond distances, when at least one of the channels is unitary.
\begin{lemma}[Average distance and diamond distance]
\label{lem:dist-avg-diam}
    Consider a unitary $U \in U(N)$ and the associated unitary channel $\mathcal{U}$, as well as a CPTP map $\mathcal{E}: \mathcal{S}_N \rightarrow \mathcal{S}_N$. Then,
    \begin{equation}
        d_{\mathrm{avg}}(\mathcal{U}, \mathcal{E}) \leq d_{\diamond}(\mathcal{U}, \mathcal{E}).
    \end{equation}
\end{lemma}
\begin{proof}
    \begin{align}
           d_{\mathrm{avg}}(\mathcal{U}, \mathcal{E})  &= \underset{|\psi\rangle \sim \mu_S}{\mathbf{E}}\left[1 - F\left(\mathcal{U}(|\psi\rangle\langle\psi|), \mathcal{E}(|\psi\rangle\langle\psi|)\right)\right]
           \\ &\leq  \underset{|\psi\rangle \sim \mu_S}{\mathbf{E}}\left[d_{\mathrm{tr}}\left(\mathcal{U}(|\psi\rangle\langle\psi|), \mathcal{E}(|\psi\rangle\langle\psi|)\right)\right]
           \\ &= \frac{1}{2}  \underset{|\psi\rangle \sim \mu_S}{\mathbf{E}}\norm{\mathcal{U}(|\psi\rangle\langle\psi|) - \mathcal{E}(|\psi\rangle\langle\psi|)}_1 
           \\&\leq \frac{1}{2} \max_{\rho \in \mathcal{S}_N}\norm{\mathcal{U}(\rho) - \mathcal{E}(\rho)}_1
           \\&\leq \frac{1}{2} \max_{\rho \in \mathcal{S}_{N^2}}\norm{(\mathcal{U} \otimes I)(\rho) - (\mathcal{E} \otimes I)(\rho)}_1
           \\& = d_{\diamond}(\mathcal{U}, \mathcal{E}),
        \end{align}
    where the first inequality uses the fact that the output of a unitary on a pure state is pure, and that when one of the states is pure, $1-F \leq d_{\mathrm{tr}}$.
\end{proof}
We also state the following elementary lower bound on the diamond distance between the depolarizing channel and any unitary channel.
\begin{lemma}[Unitaries are far from the maximally depolarizing channel]
\label{lem:diamond-distance-dep}
    For any unitary $U \in U(N)$, we have :
    \begin{equation}
        \|U(.)U^\dag - \phidep\|_\diamond \geq 2-\frac{2}{N}.
    \end{equation}
\end{lemma}
\begin{proof}
    Denote $\mathcal{U} = U (\cdot) U^\dag$. Then,
    \begin{align}
        \|\mathcal{U} - \phidep\|_\diamond &\geq \left\|\mathcal{U}(\ketbra{0}{0}) - \phidep(\ketbra{0}{0})\right\|_1
        \\&= \left\|U\ketbra{0}{0}U^\dag - \frac{I}{N}\right\|_1
        \\&= 2 - \frac{2}{N},
    \end{align}
    where the first inequality follows from Definition~\ref{def:diamond} and the fact that the norm obtained by maximizing over all input states is at least as large as that for any fixed input state.
\end{proof}

\subsection{Random Quantum Circuits}
\begin{figure}
    \centering
    \includegraphics[width = 0.4\textwidth]{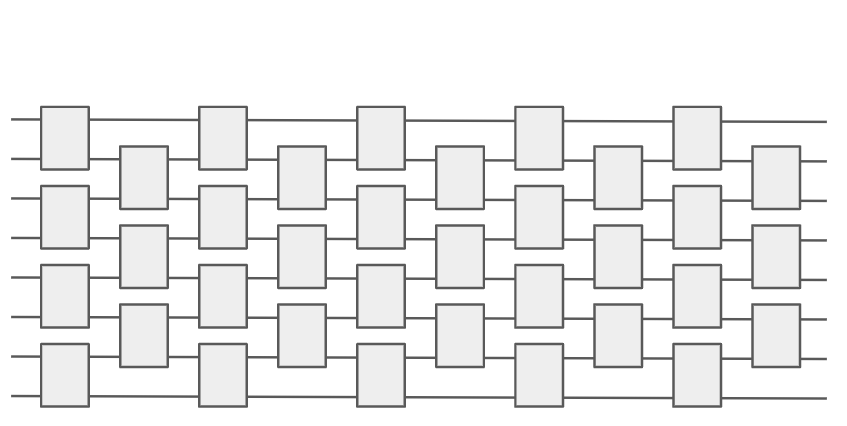}
    \caption{\small Brickwork random quantum circuits, where each gate corresponds to a 2-qubit Haar-random unitary.}
    \label{fig:brqc}
\end{figure}

We start by defining the Haar measure $\mu_H$, which can be thought of as the uniform probability distribution over $U(N)$. Similarly, we denote by $\mu_S$ the Haar measure over all pure states in $S_N$. For a comprehensive introduction to the Haar measure and its properties, we refer to \cite{mele2024introduction}.
\begin{definition}[Haar measure] 
\label{def:haar}
The Haar measure on the unitary group $U(N)$ is the unique probability measure $\mu_H$
that is both left and right invariant over $U(N)$, i.e., for all integrable functions $f$ and for all $V \in U(N)$, we
have:
\begin{equation}
\int_{U(N)} f(U) d\mu_H(U) = \int_{U(N)} f(UV) d\mu_H(U) = \int_{U(N)} f(VU) d\mu_H(U).
\end{equation}
Given a state $\ket{\phi}\in\mathbb{C}^N$, we denote the $k$-th moment of a Haar random state as
\begin{equation}
\mathbf{E}_{\ket{\psi}\sim\mu_S}\left[\ket{\psi}\bra{\psi}^{\otimes k}\right]:= \mathbf{E}_{U\sim\mu_H}\left[U^{\otimes k}\ket{\phi}\bra{\phi}^{\otimes k}U^{\dag\otimes k}\right].
\end{equation}
Note that the right invariance of the Haar measure implies that the definition of $\mathbf{E}_{\ket{\psi}\sim\mu_S}\left[\ket{\psi}\bra{\psi}^{\otimes k}\right]$ does not depend on the choice of $\ket{\phi}$.
\end{definition}
Next we will define unitary $t$-designs, which are measures with $t$-order moments matching those of the Haar measure. 
\begin{definition}[Unitary $t$-Designs]
\label{def:tdesign}
 The $t$-th moment superoperator with respect to a measure $\nu$ over $U(N)$ is defined as
    \begin{equation}
        \mathcal{M}_\nu^{(t)}(A) = \underset{U \sim \nu}{\mathbf{E}} [U^{\otimes t} A (U^\dag)^{\otimes t}] =  \int U^{\otimes t} A (U^\dag)^{\otimes t} d\nu(U).
    \end{equation}
    Then, $\nu$ is said to be an exact unitary $t$-design if and only if
    \begin{equation}
        \mathcal{M}_\nu^{(t)}(A) =  \mathcal{M}_{\mu_H}^{(t)}(A).
    \end{equation}
    Similarly, $\nu$ is said to be an additive $\delta$-approximate unitary $t$-design if and only if
    \begin{equation}
        \norm{\mathcal{M}_\nu^{(t)}(A) - \mathcal{M}_{\mu_H}^{(t)}(A)}_\diamond \leq \delta.
    \end{equation}
    We denote an exact unitary $t$-design by $\mu_H^{(t)}$ and an additive $\delta$-approximate unitary $t$-design by $\mu_H^{(t,\delta)}$.
\end{definition}
We now state some useful properties of first and second-order moments of the Haar measure.
\begin{lemma}[Moments over the Haar-measure, cf.~\cite{wadhwa2024learning,mele2024introduction}]
\label{lem:haar-moments}
    The first moment superoperator of the Haar measure $\mu_H$ over $U(N)$ is the maximally depolarizing channel.
    \begin{equation}
        \mathcal{M}_{\mu_H}^{(1)} = \phidep. 
    \end{equation}
    Further, for all $O \in H_{N,N}$ and $\rho \in \mathcal{S}_N$,
    \begin{equation}
        \underset{U \sim \mu_H}{\mathbf{E}} \left[\Tr(OU\rho U^\dag)\right] = \frac{\Tr(O)}{N}.
    \end{equation}
    The second moment is given by
    \begin{equation}
         \underset{U \sim \mu_H}{\mathbf{E}}\left[\Tr(OU \rho U^\dag)^2\right] = \left(\frac{N-\Tr(\rho^2)}{N(N^2-1)}\right)\Tr(O)^2 + \left(\frac{N\Tr(\rho^2)-1}{N(N^2-1)}\right)\Tr(O^2),
    \end{equation}
    and the variance is bounded by
    \begin{equation}
        \underset{U \sim \mu_H}{\mathbf{Var}}\left[\Tr(OU \rho U^\dag)\right] \leq \frac{1}{N+1}.
    \end{equation}
\end{lemma}
Next, we will define brickwork random quantum circuits.
\begin{definition}[Brickwork random quantum circuits]
\label{def:brqc}
    Denote by $\mathsf{RQC}(n,d)$ the measure over brickwork random quantum circuits of $n$ qubits with depth $d$. $\mathsf{RQC}(n,d)$ consists of unitaries of the form
    \begin{equation}
        U = (I_2 \otimes U_{2,3}^{(d)} \otimes U_{4,5}^{(d)} \otimes \dots) (U_{1,2}^{(d-1)} \otimes U_{3,4}^{(d-1)} \otimes \dots) \dots (U_{1,2}^{1} \otimes U_{3,4}^{1} \otimes \dots),
    \end{equation}
    where $U_{i,j}^{(l)}$ are 2-qubit unitaries distributed according to the Haar-measure over $\mathcal{U}(4)$ and $I_2$ is the identity on a single qubit (Figure~\ref{fig:brqc}). $\mathsf{RQC}(n,d)$ for any $d > 0$ forms an exact unitary 1-design. At infinite depth, the distribution over brickwork random quantum circuits converges to the Haar measure.
\end{definition}

\section{Quantum Statistical Query Oracles}
\label{sec:oracles}
In this section, we start by recalling previously defined statistical query oracles for learning quantum states and processes. We then extend these definitions naturally to define a new oracle for learning unknown observables and a new multi-copy statistical query oracle for learning quantum processes. To help keep track of all the oracles, we summarize them in Table~\ref{tab:qsq} at the end of the section.
\begin{definition}[QSQs for learning quantum states cf.~\cite{arunachalam2021quantum,arunachalam2024role,nietner2023unifying}]
\label{def:qstat}
    A quantum statistical query oracle $\mathsf{QStat}_\rho$ associated with a state $\rho \in \mathcal{S}^N$ takes as input an observable $O \in H_{N,N}$ with $\norm{O}_\infty \leq 1$ and tolerance $\tau \in \mathbb{R}, \tau > 0$, and returns $\alpha \in \mathbb{R}$ satisfying
    \begin{equation}
        |\alpha - \Tr(O\rho)| \leq \tau.
    \end{equation}
\end{definition}
\begin{definition}[Multi-copy QSQs for learning quantum states cf.~\cite{nietner2023unifying}]
\label{def:mqstat}
    A multi-copy quantum statistical query oracle $\mathsf{MQStat}_\rho^k$ associated with a state $\rho \in \mathcal{S}^N$ takes as input an observable $O \in H_{N^k,N^k}$ with $\norm{O}_\infty \leq 1$ and tolerance $\tau \in \mathbb{R}, \tau > 0$, and returns $\alpha \in \mathbb{R}$ satisfying
    \begin{equation}
        |\alpha - \Tr(O\rho^{\otimes k})| \leq \tau.
    \end{equation}
    For $k = 2$, we denote the oracle by $\mathsf{2QStat_\rho}$ instead.
\end{definition}
\begin{definition}[QSQs for learning quantum processes (QPSQs) cf.~\cite{wadhwa2024learning}]
\label{def:qpstat}
     A quantum statistical query oracle $\mathsf{QPStat}_{\mathcal{E}}$ associated with a quantum process $\mathcal{E}: \mathcal{S}_N \rightarrow \mathcal{S}_N$, takes as input an observable $O \in H_{N,N}$ with $\norm{O}_\infty \leq 1$, a state $\rho \in \mathcal{S}^N$,  and a tolerance $\tau \in \mathbb{R}, \tau > 0$ and returns $\alpha \in \mathbb{R}$ satisfying
\begin{equation}
     \\|\alpha - \Tr(O\mathcal{E}(\rho))| \leq \tau.
\end{equation}
\end{definition}
We will now define two new quantum statistical query oracles. First, we naturally extend the definition of quantum statistical queries to the multi-copy setting for quantum processes. Here, we allow the learner to query the oracle with a state that may be entangled across $k$ registers as well as with entangled measurements while applying $k$-copies of the process in parallel.
\begin{definition}[Multi-copy QPSQs]
\label{def:mqpstat}
     A multi-copy quantum statistical query oracle $\mathsf{MQPStat}_{\mathcal{E}}^k$ associated with a quantum process $\mathcal{E}: \mathcal{S}_N \rightarrow \mathcal{S}_N$, takes as input an observable $O \in H_{N^k,N^k}$ with $\norm{O}_\infty \leq 1$, a state $\rho \in \mathcal{S}^{N^k}$,  and a tolerance $\tau \in \mathbb{R}, \tau > 0$ and returns $\alpha \in \mathbb{R}$ satisfying
     \begin{equation}
     \\|\alpha - \Tr(O\mathcal{E}^{\otimes k}(\rho))| \leq \tau.
\end{equation}
Again, for $k = 2$, we denote the oracle by $\mathsf{2QPStat_\mathcal{E}}$ instead.
\end{definition}
Next, we present a new definition of quantum statistical queries for learning unknown observables. 
\begin{definition}[QSQs for learning observables]
\label{def:qstato}
    A quantum statistical query oracle $\mathsf{QStat}_O$ associated with an observable $O \in H_{N,N}$ with $\norm{O}_\infty \leq 1$, takes as input a state $\rho \in \mathcal{S}^N$ and tolerance $\tau \in \mathbb{R}, \tau > 0$, and returns $\alpha \in \mathbb{R}$ satisfying
    \begin{equation}
        |\alpha - \Tr(O\rho)| \leq \tau.
    \end{equation}
\end{definition}
We summarize these definitions in Table \ref{tab:qsq} below.
\begin{table}[h!]
    \centering
    \begin{tabular}{|c|c|c|c|c|}
         \hline
         Oracle & Object to learn & Inputs & Output($ \pm\tau$) \\
         \hline
         $\mathsf{QStat}_\rho$ & $\rho \in \mathcal{S}_N$ & $O \in H_{N,N}, \norm{O}_\infty \leq 1$ & $\Tr(O\rho)$ \\
         $\mathsf{MQStat}_\rho^k$ & $\rho \in \mathcal{S}_N$ & $O \in H_{N^k,N^k}, \norm{O}_\infty \leq 1$ & $\Tr(O\rho^{\otimes k})$ \\
         $\mathsf{QPStat}_\mathcal{E}$ & $\mathcal{E}: \mathcal{S}_N \rightarrow \mathcal{S}_N$ & $\rho \in \mathcal{S}_N, O \in H_{N,N}, \norm{O}_\infty \leq 1$ & $\Tr(O\mathcal{E}(\rho))$
         \\
         $\mathsf{MQPStat}_\mathcal{E}^k$ [\textit{\small This work}] & $\mathcal{E}: \mathcal{S}_N \rightarrow \mathcal{S}_N$ & $\rho \in \mathcal{S}_{N^k}, O \in H_{N^k,N^k}, \norm{O}_\infty \leq 1$ & $\Tr(O\mathcal{E}^{\otimes k}(\rho))$
         \\
         $\mathsf{QStat}_O$ [\textit{\small This work}] & $O \in H_{N,N}, \norm{O}_\infty \leq 1$ & $\rho \in \mathcal{S}_N$  & $\Tr(O\rho)$ \\
         \hline
    \end{tabular}
    \caption{Summary of quantum statistical query oracles}
    \label{tab:qsq}
\end{table}

\section{Noise-tolerance of QPSQ}
\label{sec:robust}
Statistical query algorithms, in classical learning theory, are known to be robust to classification noise~\cite{kearns1998efficient}. A similar result was shown for quantum statistical queries in~\cite{gollakota2022hardness} for classification noise, depolarizing noise and any bounded noise. We now demonstrate the noise-robustness of learners with access to the \textsf{QPStat} oracle. Informally, we show that a QPSQ learning algorithm for a class of quantum channels in a noiseless setting can successfully learn the same class even when given noisy data; more specifically, with \textsf{QPStat} access to the noisy version of the channel (within a noise threshold that is not too high). In particular, the number of queries needed and the algorithm itself remain unchanged, while the tolerance of the queries needs to be lowered. We formalize this result in the following theorem.
\begin{theorem}[Noise tolerance of QPSQ learner]
\label{thm:qpsq-robust}
    Suppose there exists a learning algorithm that learns a class of channels $ \mathcal{C} = \{\mathcal{E}_i\}_i$ using $q$ queries to the oracle $\mathsf{QPStat}_{\mathcal{E}}$ for an unknown channel $\mathcal{E} \in \mathcal{C}$, with tolerance at least $\tau$, for observables $\{O_j\}_{j \in [q]},$ with $\norm{O_j}_\infty \leq 1$, $\forall j \in [q]$. Let $\Lambda$ be any unknown noise channel with the guarantee that for some $\eta, 0 < \eta < \tau$, and for all $\mathcal{E}_i \in \mathcal{C}$, we have
    \begin{equation}
        \norm{\Lambda(\mathcal{E}_i) - \mathcal{E}_i}_\diamond \leq \eta.
    \end{equation}
    Then, there exists an algorithm for learning $\mathcal{C}$ given access to $\mathsf{QPStat}_{\Lambda(\mathcal{E}_i)}$, using $q$ queries of tolerance at least $\tau-\eta$.
\end{theorem}
\begin{proof}
Let the algorithm in the noisy scenario make the exact same queries as in the noiseless case while changing the tolerance from $\tau$ to $\tau-\eta$. Then, on input $O,\rho$, the oracle responds with $\alpha$ such that 
\begin{equation}
    |\alpha - \Tr(O\Lambda(\mathcal{E}_i)(\rho))| \leq \tau - \eta.    
\end{equation}
By the definition of $\|\cdot\|_\diamond$ and the matrix H\"{o}lder inequality, we have
    \begin{equation}
        |\Tr(O\mathcal{E}(\rho))| \leq \norm{O}_\infty \norm{\rho}_{1} \norm{\mathcal{E}}_\diamond \leq \norm{\mathcal{E}}_\diamond.
    \end{equation}
    Therefore, 
    \begin{equation}
        |\Tr(O\mathcal{E}_i(\rho)) - \Tr(O\Lambda(\mathcal{E}_i)(\rho))| \leq \norm{\mathcal{E}_i-\Lambda(\mathcal{E}_i)}_\diamond \leq \eta.
    \end{equation}
    From the triangle inequality, we have
    \begin{equation}
        |\alpha - \Tr(O\mathcal{E}_i(\rho))| \leq \tau .
    \end{equation}
    Thus, the learner in the noisy setting now receives data with identical guarantees to the noiseless case and can proceed in the same way.
\end{proof}
\subsection{Estimating noise with a single query to \textsf{2QPStat}}
As Theorem~\ref{thm:qpsq-robust} requires us to reduce the query tolerance by the value of the diamond norm between the noisy and noiseless channels, it is important to identify this quantity. In many situations, such an upper bound may already be known from previous benchmarking. However, it is important to consider whether such a bound can also be obtained using \emph{only} statistical queries. We show a method to perform such an estimation when the noise is global depolarizing in Theorem~\ref{thm:qpsq-noise-estim}.

However, an immediate challenge to this problem is that the strength of the depolarizing noise is closely related to the purity of the output state, which has been shown to be hard to estimate from single-copy quantum statistical queries~\cite{arunachalam2024role}. Nevertheless, it was shown in~\cite{nietner2023unifying} that a single query to $\mathsf{2QStat}$ suffices to estimate the purity of a state. We use this single-query purity estimation method to characterize the depolarizing strength from a single query to $\mathsf{2QPStat}$. For our estimation to succeed, we require prior knowledge of an upper bound on the depolarizing noise. However, this bound need not be tight. For example, $\gamma \leq 1/2$ is a good enough bound for our purposes and would hold in most practical situations.
\begin{theorem}[Estimating noise with a single $\mathsf{2QPStat}$ query]
    \label{thm:qpsq-noise-estim}
    Consider a noiseless unitary channel $\mathcal{U} = U (.) U^\dag$, and the corresponding noisy channel $\Lambda(\mathcal{U}) = \Lambda(\gamma) \circ \mathcal{U}$, where $\Lambda(\gamma)$ is the depolarizing channel with strength $\gamma$:
    \begin{equation}
        \Lambda({\gamma}) : \rho \rightarrow (1-\gamma)\rho + \gamma \Tr(\rho) \frac{\mathbb{I}}{d}.
    \end{equation}Then, given an initial (loose) upper bound  $\gamma \leq \gamma_u$, there exists a method to characterize the noise, 
    \begin{equation}
        \norm{\mathcal{U}-\Lambda(\mathcal{U})}_\diamond \in [l,u],
    \end{equation}
    such that $u - l \leq \epsilon$, for $0 < \epsilon \leq 1-\gamma_u$, using a single query to $\mathsf{2QPStat_{\Lambda(\mathcal{U})}}$ with tolerance $\tau = \Theta((1-\gamma_u)\epsilon)$.
\end{theorem}
\begin{proof}[Proof sketch]
    We will use a single query to $\mathsf{2QPStat}$ to estimate the purity of the output state. First, observe that 
\begin{equation}
    \Tr(\mathbb{F}(\Lambda(\mathcal{U})(|\psi\rangle\langle\psi|))^{\otimes 2}) = \Tr((\Lambda(\mathcal{U})(|\psi\rangle\langle\psi|))^2),
\end{equation}
giving us the purity of the output state, where $\mathbb{F}$ is the flip operator. A $\tau$-accurate estimate of this quantity can be obtained by making a query of the form $\mathsf{2QPStat}_{\Lambda(\mathcal{U})}(|\psi\rangle\langle\psi|^{\otimes 2}, \mathbb{F}, \tau)$. Note that the purity of the output state will be the same for any pure input $|\psi\rangle$. For simplicity, we choose $|\psi\rangle = |0\rangle$. We define the output state
\begin{equation}
    \rho^{\mathrm{out}} = (1-\gamma)U|0\rangle\langle0|U^\dag + \gamma \frac{\mathbb{I}}{2^n}.
\end{equation}
We have 
\begin{equation}
        (\rho^{\mathrm{out}})^2 =  (1-\gamma)^2 U|0\rangle\langle0|U^\dag + \frac{2\gamma(1-\gamma)}{2^n}U|0\rangle\langle0|U^\dag + \gamma^2 \frac{\mathbb{I}}{4^n}.
\end{equation}
This gives us the purity of the output state
\begin{align}
        \Tr((\rho^{out})^2) &= (1-\gamma)^2 + \frac{2\gamma(1-\gamma)}{2^n} + \frac{\gamma^2}{2^n} \\
        &= 1 - (2\gamma-\gamma^2)(1-\frac{1}{2^n}).
\end{align}
Suppose the query to \textsf{2QPStat} gives us a quantity $\alpha$, 
\begin{equation}
    \alpha \leftarrow \mathsf{2QPStat}_{\Lambda(\mathcal{U})}(|0\rangle\langle0|^{\otimes 2}, \mathbb{F}, \tau).
\end{equation}
Then, by definition of the oracle,
\begin{equation}
    \alpha \in \left[ 1-(2\gamma-\gamma^2)(1-2^{-n})-\tau, 1-(2\gamma-\gamma^2)(1-2^{-n})+\tau\right]
\end{equation}
We will use this range of the estimated purity to find a range for $\gamma$, and then extend it to characterize the diamond distance between the noiseless and noisy channels. We defer the rest of the analysis to Appendix~\ref{app:noise-estimation}.
\end{proof}
In practice, we are interested in the regime when $\tau \geq 1/\mathsf{poly}(n)$, and thus $\epsilon \geq 1/\mathsf{poly}(n)$ and $1-\gamma_u \geq 1/\mathsf{poly}(n)$. For known upper bounds on the depolarizing strength that are at least inverse polynomially bounded from $1$, one can thus obtain an estimate of the diamond norm between the noisy and noiseless channels up to inverse-polynomial precision using a single query to \textsf{2QPStat} of inverse-polynomial tolerance. 
\\\newline As stated earlier, this method only applies to global depolarizing noise. While this is a physically relevant noise model, it would be interesting to consider the estimation of more general kinds of noise from statistical queries.

\section{QPSQ learner for shallow circuits}
\label{sec:learner}
In this section, we show an efficient algorithm for learning constant-depth circuits within diamond distance.
Huang \textit{et al}.~\cite{huang2024learning} showed an efficient learning algorithm for this problem using classical shadows. The algorithm proceeds by learning all $3n$ single-qubit Pauli observables after Heisenberg-evolution under the unknown circuit, using classical shadows of the circuit, similar to the algorithm of~\cite{huang2023learning}. Then, the algorithm combines these learned observables using a novel \textit{circuit sewing} procedure. The algorithm only makes random queries to the circuit to learn the observables, and has a sample complexity of $\mathcal{O}\left(n^2 \log (n/\delta)/\epsilon^2\right)$. We show that it is possible to learn all the Heisenberg-evolved Pauli observables using $\mathcal{O}\left(n^3 \log (n/\delta)/\epsilon^2\right)$ queries to $\mathsf{QPStat}_\mathcal{U}$. This allows us to efficiently learn constant-depth circuits using statistical queries within bounded diamond distance. Thus, with a linear overhead in the query complexity, we can gain the robustness guarantees of Theorem~\ref{thm:qpsq-robust} for learning any constant-depth circuit. 

We upper bound the query complexity of our algorithm in the following theorem.
\begin{theorem}[Learning  quantum circuits from \textsf{QPStat} queries]
\label{thm:qpsq-shallow-learner}
    There exists a QPSQ algorithm for learning an unknown $n$-qubit unitary $U$ generated by a constant-depth circuit over any two-qubit gates in $\mathrm{SU}(4)$, and with an arbitrary number of ancilla qubits, such that the algorithm outputs an $n$-qubit quantum channel $\hat{\mathcal{E}}$ that can be implemented by a constant-depth quantum circuit over 2n qubits, which satisfies 
    \begin{equation}
        \norm{\hat{\mathcal{E}} - \mathcal{U}}_\diamond \leq \epsilon
    \end{equation}
    with probability at least $1-\delta$. The algorithm uses 
    \begin{equation}
        N = \mathcal{O}\left(\frac{n^3 \log(n/\delta)}{\epsilon^2}\right)
    \end{equation}
    queries to $\mathsf{QPStat}_{\mathcal{U}}$ with tolerance $\tau = \Omega(\epsilon/n)$ and runs in computational time $\mathcal{O}(\mathsf{poly}(n)\log(1/\delta)/\epsilon^2)$
\end{theorem}
The rest of this section is devoted to proving Theorem \ref{thm:qpsq-shallow-learner}. First, we state a key result from \cite{huang2024learning}, showing that learning Heisenberg-evolved single-qubit Paulis suffices to learn a unitary.
\begin{lemma}[Circuit sewing cf.~\cite{huang2024learning}]
\label{lem:circuit-sewing}
    Let $U \in U(2^n)$. Let $O_{i,P} = U^\dag P_i U$ be the $3n$ Heisenberg-evolved single-qubit Pauli observables, where $P \in \{X,Y,Z\}, i \in [n]$. Then, given descriptions of operators $\hat{O}_{i,P}$ satisfying
    \begin{equation}
        \|O_{i,P} - \hat{O}_{i,P}\| \leq \epsilon/(6n),
    \end{equation}
    for all $i \in [n],  P \in \{X,Y,Z\}$, one can construct a channel $\hat{\mathcal{E}}$ such that
    \begin{equation}
        \|\hat{\mathcal{E}} - U(\cdot)U^\dag\|_\diamond \leq \epsilon,
    \end{equation}
    in $\poly(n)\log(1/\delta)/\epsilon^2$ computational time.
\end{lemma}

Next, we will show a quantum statistical query algorithm for learning a few-body observable with unknown support. We denote the support of an observable O, i.e. the set of qubits it acts on, by $\mathsf{supp}(O)$. 
\begin{lemma}[Learning a few-body observable with unknown support from QSQs] 
\label{lem:learn-observable}
There exists a QSQ algorithm for learning an unknown $n$-qubit observable $O$, with $\norm{O}_\infty \leq 1,$ that acts on an unknown set of $k$ qubits such that with probability at least $1-\delta$, the learned observable $\hat{O}$ satisfies
\begin{equation}
    \norm{\hat{O}-O} \leq \epsilon \quad \mathrm{and} \quad  \mathsf{supp}(\hat{O}) \subseteq \mathsf{supp}(O),
\end{equation}
using 
\begin{equation}
\label{eq:queries-unknown-observable}
    N = \frac{2^{\mathcal{O}(k)} \log(n/\delta)}{\epsilon^2}
\end{equation}
queries to $\mathsf{QStat}_O$ of tolerance 
\begin{equation}
\label{eq:tolerance-unknown-observable}
    \tau = \frac{\epsilon}{4\left(6\sqrt{2}\right)^k},
\end{equation}
running in computational time $\mathcal{O}(n^k \log(n/\delta)/\epsilon^2)$.
\end{lemma}
\begin{proof}
    Consider $O = \underset{{P \in \mathcal{P}_n, |P| \leq k}}{\sum} \alpha_P P$. The Pauli coefficients $\alpha_P$ can be represented as
    \begin{equation}
        \alpha_P = 3^{|P|} \underset{|\psi\rangle \sim \mathrm{stab}_1^{\otimes n}}{\mathbf{E}} \langle \psi | O | \psi \rangle \langle \psi | P | \psi \rangle.
    \end{equation}
    The algorithm makes random queries and uses the output to estimate each coefficient. Specifically, it makes $N$ queries $\mathsf{QStat}_{O}(|\psi_l\rangle\langle\psi_l|, \tau)$ for $l \in [N]$, with random input states $|\psi_l\rangle \sim \mathrm{stab}_1^{\otimes n}$. Denote the output of the $l^{\mathrm{th}}$ query by $y_l$.
    Denote by $\hat{\alpha}_P$ the estimate
    \begin{equation}
        \hat{\alpha}_P = \frac{3^{|P|}}{N} \sum_{l \in [N]} y_l  \langle \psi_l | P | \psi_l \rangle.
    \end{equation}
    Define the coefficients of the learned observables as
    \begin{equation}
        \hat{\beta}_P = 
        \begin{cases}
            \hat{\alpha}_P &  \hat{\alpha}_P > 0.5\epsilon/(2\sqrt{2})^k,\\
            0 &           \hat{\alpha}_P \leq 0.5\epsilon/(2\sqrt{2})^k.  
        \end{cases}
    \end{equation}
    Then, the algorithm outputs the observable 
    \begin{equation}
        \hat{O} = \underset{{P \in \mathcal{P}_n, |P| \leq k}}{\sum} \hat{\beta}_P P.
    \end{equation}
    First, we show the correctness of the algorithm. Denote the intermediate quantity $\Bar{\alpha}_P$
    \begin{equation}
        \Bar{\alpha}_P = \frac{3^{|P|}}{N} \sum_{l \in [N]} \langle \psi_l | O | \psi_l \rangle  \langle \psi_l | P | \psi_l \rangle.
    \end{equation}
    By the definition of $\mathsf{QStat}_O$,
    \begin{equation}
        |\hat{\alpha}_P - \Bar{\alpha}_P| \leq 3^{|P|}\tau.
    \end{equation}
    Using $N$ queries as specified in (\ref{eq:queries-unknown-observable}), with the tolerance $\tau$ given in (\ref{eq:tolerance-unknown-observable}), from Hoeffding's inequality, we see that with probability at least $1-\delta$, it holds for all $P \in \mathcal{P}_n, |P| \leq k$,
    \begin{equation}
        |\Bar{\alpha}_P - \alpha_P| \leq \frac{\epsilon}{4(2\sqrt{2})^k} \leq  \frac{0.5\epsilon}{(2\sqrt{2})^k} - 3^{k}\tau \leq \frac{0.5\epsilon}{(2\sqrt{2})^k} - 3^{|P|}\tau.
    \end{equation}
    Thus, $N$ queries of tolerance $\tau$ suffice to obtain, for all $P \in \mathcal{P}_n, |P| \leq k$, 
    \begin{equation}
    \label{eq:alpha-estimate}
        |\hat{\alpha}_P - \alpha_P| \leq \frac{0.5\epsilon}{(2\sqrt{2})^k}.
    \end{equation}
    From this point, we follow the presentation of~\cite{huang2024learning}. First, we show that $\mathsf{supp}(\hat{O}) \subseteq \mathsf{supp}(O)$. For all $P \in \mathcal{P}_n$ with $\alpha_P = 0$, (\ref{eq:alpha-estimate}) tells us $|\hat{\alpha}_P| \leq \frac{0.5\epsilon}{(2\sqrt{2})^k}$. Thus, $\hat{\beta}_P = 0$, showing that $\mathsf{supp}(\hat{O}) \subseteq \mathsf{supp}(O)$. \\\newline
    Now, we prove the error bound on the learned observable. As $\alpha_P = 0$ implies $\hat{\beta}_P = 0$, we have
    \begin{align}
            \hat{O}-O &= \underset{P \in \mathcal{P}_n : \mathsf{supp}(P) \subseteq \mathsf{supp}(O)}{\sum} \left( \hat{\beta}_P - \alpha_P \right)P \\&= \sum_{Q \in \mathcal{P}_k} \left( \hat{\beta}_{P(Q)} - \alpha_{P(Q)}\right) P(Q),
    \end{align}
    where $P(Q)$ denotes $Q \otimes I_{[n] \backslash \mathsf{supp}(O)}$, where $I_{[n] \backslash \mathsf{supp}(O)}$ is the identity on all qubits outside the support of $O$. Now,
    \begin{align}
        \norm{\hat{O}-O}_\infty &= \norm{\sum_{Q \in \mathcal{P}_k} \left( \hat{\beta}_{P(Q)} - \alpha_{P(Q)}\right) P(Q)}_\infty \\&= \norm{\sum_{Q \in \mathcal{P}_k} \left( \hat{\beta}_{P(Q)} - \alpha_{P(Q)}\right) Q}_\infty \\&\leq \sqrt{\sum_{Q \in \mathcal{P}_k} \left( \hat{\beta}_{P(Q)} - \alpha_{P(Q)}\right)^2 \Tr(Q^2)} \\&\leq (2\sqrt{2})^k \max_{|P| \leq k}\left| \hat{\beta}_P - \alpha_P\right| \\&\leq (2\sqrt{2})^k \max_{|P| \leq k}\left(\left| \hat{\beta}_P - \hat{\alpha}_P\right| + \left| \hat{\alpha}_P - \alpha_P\right|\right)
        \\& \leq \epsilon,
    \end{align}
    where the first inequality follows from the fact that $\norm{A}_\infty \leq \sqrt{\Tr(A^2)}$ for any Hermitian matrix $A$, the second inequality follows from the facts that Q is a $k$-qubit Pauli and that there are $4^k$ terms in the summation, and the third inequality follows from triangle inequality.
\end{proof}
We will also use the following lemma from~\cite{huang2024learning} on the size of the support of single-qubit Paulis after Heisenberg evolution under constant depth circuits.
\begin{lemma}[Support of Heisenberg-evolved observables is bounded cf.~\cite{huang2024learning}, Lemma 14]
    \label{lem:bounded-support}
    Given an $n$-qubit unitary $U$ generated by a constant-depth circuit. For each qubit $i \in [n]$ and Pauli operator $P \in \{X,Y,Z\}$, we have
    \begin{equation}
        |\mathsf{supp}(U^\dag P_i U)| = \mathcal{O}(1).
    \end{equation}
\end{lemma}
We can now prove Theorem \ref{thm:qpsq-shallow-learner}.
\begin{proof}[Proof of Theorem \ref{thm:qpsq-shallow-learner}]
    Our algorithm uses $\mathsf{QPStat}_{U}$ queries to learn the $3n$ Heisenberg-evolved observables $O_{i,P} = U^\dag P_i U, P \in \{X,Y,Z\}, i \in [n]$. 
    Note that 
    \begin{equation}
        \mathsf{QPStat}_{U}(\rho, P_i, \tau) \equiv \mathsf{QStat}_{O_{i,P}}(\rho,\tau),
    \end{equation}
    i.e. one can make statistical queries to $O_{i,P}$ using statistical queries to the unitary $U$.
    Thus, one can run the algorithm of Lemma~\ref{lem:learn-observable} for learning few-body observables by using queries to $\mathsf{QPStat}_U$. From Lemma \ref{lem:circuit-sewing}, we see that to learn the unitary successfully, we require the algorithm of Lemma~\ref{lem:learn-observable} to learn each $O_{i,P}$ independently up to error $\epsilon/6n$ with probability at least $1-\delta/3n$. From Lemma \ref{lem:bounded-support}, we see that all observables have support $\mathcal{O}(1)$. Thus, from Lemma \ref{lem:learn-observable}, we obtain the desired query complexity $N = \mathcal{O}\left(\frac{n^3 \log(n/\delta)}{\epsilon^2}\right)$ and tolerance $\tau = \Omega(\epsilon/n)$. The overall computational time is $\mathcal{O}(\mathsf{poly}(n)\log(1/\delta)/\epsilon^2)$, and can be obtained from Lemmas \ref{lem:circuit-sewing} and~\ref{lem:learn-observable}.
\end{proof}

\section{Lower bounds for shallow random quantum circuits}
\label{sec:hardness}
In this section, we will show an average-case query complexity lower bound for learning brickwork random quantum circuits (See Definition \ref{def:brqc}) with depth at least logarithmic and at most linear in the number of qubits. We formalize this result in the following theorem.
\begin{theorem}[Average-case lower bound for shallow BRQCs]
\label{thm:qpsq-diamond-depth}
    Let $N = 2^n, 0 < \tau \leq \epsilon \leq \frac{1}{3}\left(1-\frac{1}{N}\right)$, $\mathsf{RQC}(n,d)$ be an ensemble of $n$-qubit brickwork-random quantum circuits of depth $d$, with
    \begin{equation}
      \frac{\log (n)}{\log (5/4)} \leq d \leq \frac{n+ \log (n)}{\log (5/4)}.
    \end{equation}Assume there exists an algorithm that with probability $\beta$ over $U \sim \mathsf{RQC}(n,d)$  and probability $\alpha$ over its internal randomness produces a hypothesis $\hat{\mathcal{E}}$ such that $d_\diamond(\hat{\mathcal{E}}, U(.)U^\dag)\leq \epsilon$, using $q$ queries with tolerance $\tau$ to $\mathsf{QPStat}_U$. Then, it holds 
    \begin{equation}
            q + 1 \geq \Omega\left( \frac{(2\alpha - 1)\beta}{n} \tau^2 \left(\frac{5}{4}\right)^d\right). 
    \end{equation}
\end{theorem}
Before proving this theorem, we will state some useful lemmas. We start by stating the general lower bound for learning unitaries up to diamond distance using QPSQs from~\cite{wadhwa2024learning}.
\begin{lemma}[General lower bound for QPSQ-learning within $d_\diamond$, cf.~\cite{wadhwa2024learning}]
\label{lem:qpsq-diamond-general}
     Let $0 < \tau \leq \epsilon$, $\mathcal{C} \subseteq U(N)$ be a set of unitaries, and $\mu$ some measure over $\mathcal{C}$. Assume there exists an algorithm that, with probability $\beta$ over $U \sim \mu$  and probability $\alpha$ over its internal randomness, produces a hypothesis quantum channel $\Phi$ such that $d_\diamond(\Phi, U (\cdot) U^\dag)\leq \epsilon$, using $q$ queries with tolerance $\tau$ to $\mathsf{QPStat}_{U}$. Then, $q$ must satisfy 
    \begin{equation}
        q+1 \geq \frac{(2\alpha - 1)\beta} {\underset{\rho, O}{\max\text{ }} \mathbf{Pr}_{\mathcal{E} \sim \mu} (|\Tr(O\mathcal{E}(\rho)) - \Tr(O\phidep(\rho))| > \tau)}.
    \end{equation}
\end{lemma}
To prove Theorem~\ref{thm:qpsq-diamond-depth}, we will simply upper bound the probability in the denominator of Lemma~\ref{lem:qpsq-diamond-general}. Recall that brickwork random quantum circuits of any non-zero depth form $1$-designs, and thus from Lemma \ref{lem:haar-moments}, their first moment superoperator is $\phidep$. As a result, it will suffice to bound the variance of $\Tr(OU\rho U^\dag)$ for $U \sim \mathsf{RQC}(n,d)$ to upper bound the aforementioned probability.
\begin{lemma}[Bounded variance for $\mathsf{RQC}(n,d)$]
    \label{lem:shallow-variance}
    For depth
    \begin{equation}
      \frac{\log (n)}{\log (5/4)} \leq d \leq \frac{n+ \log (n)}{\log (5/4)},
    \end{equation}
    the variance over $\mathsf{RQC}(n,d)$ is bounded by
    \begin{equation}
    \underset{U \sim \mathsf{RQC(n,d)}}{\mathbf{Var}} \left[\Tr(OU\rho U^\dag)\right] = \mathcal{O}\left( n \left(\frac{4}{5}\right)^d\right),
    \end{equation}
     for all $O \in H_{2^n,2^n}$ with $\norm{O}_\infty \leq  1$ and all $\rho \in \mathcal{S}_{2^n}$.
\end{lemma}
\begin{proof}[Proof sketch]
    Bounds on low-order moments of BRQCs of the considered depths have been shown in~\cite{nietner2023average, barak2020spoofing} by counting partitions over a statistical mechanical model, a technique developed originally in~\cite{hunter2019unitary}. Using similar arguments, we obtain the following bound.
    \begin{equation}
        \underset{U \sim \mathsf{RQC}(n,d)}{\mathbf{E}} \left[\Tr(OU\rho U^\dag)^2\right] \leq \left(1+\left(\frac{4}{5}\right)^d\right)^{n/2} \underset{U \sim \mu_H}{\mathbf{E}}\left[\Tr(OU\rho U^\dag)^2\right].
    \end{equation}
    Then, using the fact that brickwork random quantum circuits form exact 1-designs at any depth, and substituting in the moments from Lemma~\ref{lem:haar-moments}, we obtain the desired result. We defer the complete proof to Appendix~\ref{app:shallow-var}.
\end{proof}

We can now prove Theorem~\ref{thm:qpsq-diamond-depth}.
\begin{proof}[Proof of Theorem~\ref{thm:qpsq-diamond-depth}]
      As $\mathsf{RQC}(n,d)$ is a 1-design, the expected channel over this ensemble is the maximally depolarizing channel. Thus, using Chebyshev's inequality, we have
    \begin{align}
        \underset{U \sim \mathsf{RQC}(n,d)}{\mathbf{Pr}} (|\Tr(OU(\rho)U^{\dag}) - \Tr(O\phidep(\rho))| > \tau) &\leq \frac{\mathbf{Var}_{U \sim \mathsf{RQC}(n,d)}[\Tr(OU\rho U^\dag)]}{\tau^2} \\& = \mathcal{O} \left(\frac{n}{\tau^2}\left(\frac{4}{5}\right)^d\right),
    \end{align}
    where we use Lemma~\ref{lem:shallow-variance} in the second step.
    Now, using Lemma~\ref{lem:qpsq-diamond-general}, we obtain the desired result
    \begin{equation}
        q + 1 \geq \Omega\left(\frac{(2\alpha - 1)\beta}{n} \tau^2 \left(\frac{5}{4}\right)^d \right).
    \end{equation}
\end{proof}

\section{PRUs cannot be shallow}
\label{sec:pru}
In this section, we will show that constant-depth circuits cannot be used to construct pseudorandom unitaries (PRUs). We start by recalling the definition of PRUs.
\begin{definition}[Pseudorandom unitaries (PRUs) cf.~\cite{ji2018pseudorandom,metger2024simple}]
    Let $n \in \mathbb{N}$ be the security parameter. Let $\mathcal{K}$ denote the key space. An infinite sequence $\mathcal{U} = \{\mathcal{U}_n\}_{n \in \mathbf{N}}$ of $n$-qubit unitary ensembles $\mathcal{U}_n = \{U_k\}_{k \in \mathcal{K}}$ is said to be pseudorandom if
    \begin{itemize}
        \item \textbf{(Efficient computation)} There exists a polynomial-time quantum algorithm $\mathcal{Q}$ such that for all keys $k \in \mathcal{K}$, and any n-qubit pure state $|\psi\rangle$, $\mathcal{Q}(k, |\psi\rangle) = U_k |\psi\rangle$.
        \item \textbf{Pseudorandomness} The unitary $U_k$, for a random key $k \sim \mathcal{K}$, is computationally indistinguishable from a Haar-random unitary $U \sim \mu_H$. In other words, for any quantum polynomial-time algorithm $\mathcal{A}$, it holds that
        \begin{equation}
            \left|\underset{k \sim \mathcal{K}}{\mathbf{Pr}}\left[\mathcal{A}^{U_k}(1^n) = 1\right] - \underset{U \sim \mu_H}{\mathbf{Pr}}\left[\mathcal{A}^{U}(1^n) = 1\right]\right| \leq \mathrm{negl}(n).
        \end{equation}
    \end{itemize}
\end{definition}
Now, we can state the main theorem of this section, a no-go result for constructing PRUs using constant-depth circuits.
\begin{theorem}[Constant-depth unitaries cannot form PRUs]
\label{thm:pru}
    For sufficiently large $n$, a unitary sampled from any ensemble $\mathcal{C}_n$ over $n$-qubit circuits composed of 2-qubit gates with depth $\mathcal{O}(1)$ can be distinguished from a random n-qubit unitary from the Haar measure with non-negligible advantage using $\mathcal{O}\left(n^4 \log(n)\right)$ queries to the unknown unitary in time $\mathrm{poly}(n)$. Thus, $\mathcal{C}_n$ is not an ensemble of pseudorandom unitaries.
\end{theorem}

To show that constant-depth unitaries cannot form PRUs, we will construct an efficient distinguisher that achieves non-negligible advantage in distinguishing any constant-depth unitary from Haar-random unitaries. As an overview, our distinguisher will consist of the learning and verification algorithms for constant-depth circuits of~\cite{huang2024learning}. We will use these algorithms in a black-box manner. Given as input any constant-depth circuit, the learning algorithm correctly learns it with high probability. Then, the verification algorithm also passes with high probability. On the other hand, given a Haar-random unitary, the algorithm is not likely to perform well. Rather than directly analyzing the performance of the algorithm on Haar-random unitaries, we formalize this notion by extending the quantum no-free lunch theorem~\cite{poland2020no}, proving that any algorithm that learns a Haar-random unitary with high probability requires an exponential number of queries. Then, since the learned unitary is not close to the actual one, the verification algorithm fails with high probability.

We start by recalling the results of the learning and verification algorithms from~\cite{huang2024learning} in the following lemma.

\begin{lemma}[Learning and Verification Algorithms for Constant-Depth Circuits, cf.~\cite{huang2024learning}]
There exists a learning algorithm $\mathcal{A}_L(n, \epsilon, \delta)$ and a verification algorithm $\mathcal{A}_V(n,\epsilon,\delta)$ such that
\begin{itemize}
    \item (Learning) $\mathcal{A}_{L}(n, \epsilon, \delta)$ makes $\mathcal{O}(n^2\log (n/\delta)/\epsilon^2)$ queries to an $n$-qubit unitary $U$ implemented by a constant-depth circuit composed of two-qubit gates, runs in time $\mathsf{poly}(n)/\epsilon^2$, and outputs a channel $\mathcal{E}$ such that $d_\diamond(U(.)U^\dag, \mathcal{E}) \leq \epsilon$ with probability at least $1-\delta$.
    \item (Verification) Given a learned implementation of a $n$-qubit CPTP map $\hat{\mathcal{E}}$ and query access to an unknown CPTP map $\mathcal{C}$, $\mathcal{A}_V(n,\epsilon,\delta)$ makes $\mathcal{O}((n^2\log (n/\delta)/\epsilon^2)$ queries to $\mathcal{C}$, runs in computational time $\mathcal{O}((n^3\log (n/\delta)/\epsilon^2)$, and
    \begin{enumerate}
        \item If $d_{\mathrm{avg}}(\hat{\mathcal{E}}, \mathcal{C}) > \epsilon$, $\mathcal{A}_V$ outputs \textbf{FAIL} with probability at least $1-\delta$.
        \item If $d_{\mathrm{avg}}(\hat{\mathcal{E}}, \mathcal{C}) \leq \epsilon/12n$ and $d_\diamond(\mathcal{C}^\dag\mathcal{C} - \mathcal{I}) \leq \epsilon/24n$, $\mathcal{A}_V$ outputs \textbf{PASS} with probability at least $1-\delta$.
    \end{enumerate}
    Moreover, the queries made by both $\mathcal{A}_L$ and $\mathcal{A}_V$ are efficiently preparable pure states.
\end{itemize}
\end{lemma}
In other words, $\mathcal{A}_L$ efficiently learns any constant depth quantum circuit, and $\mathcal{A}_V$ ensures that the unknown channel is unitary and that it is learned correctly.
Next, we state a theorem on the average-case hardness of learning Haar-random unitaries from black-box queries, which will later allow us to prove that the distinguisher behaves as desired on Haar-random unitaries.
\begin{theorem}[Average-case hardness of learning Haar-random unitaries]
\label{thm:haar-bb-avg-lb}
    Let $\epsilon,\delta \in (0,1)$. Suppose there exists a learning algorithm that queries a unitary $U$ distributed according to the Haar-measure over $U(2^n)$ with $q$ pure states, and outputs a channel $\mathcal{E}$, such that with probability at least $1-\delta$, the channel approximates the Haar-random unitary within average distance $\epsilon$, i.e
    \begin{equation}
        \underset{U \sim \mu_H}{\mathbf{Pr}}[d_{\mathrm{avg}}(U(.)U^\dag, \mathcal{E}) \leq \epsilon] \geq 1-\delta.
    \end{equation} Then, the algorithm must make at least
    \begin{equation}
        q \geq 2^n(1-\delta)(1-\epsilon)-1
    \end{equation} queries.
\end{theorem}
We defer the proof of Theorem~\ref{thm:haar-bb-avg-lb} to Appendix~\ref{sec:haardness}. While our proof strategy results in a very loose dependence on $\epsilon$ and $\delta$, this bound is sufficient for our purposes. We can now prove Theorem~\ref{thm:pru}.
\begin{proof}[Proof of Theorem \ref{thm:pru}]
    We consider our distinguisher $\mathcal{A}$ to be the composition of the two algorithms $\mathcal{A}_L(n,\frac{1}{48n}, 1/6)$ and $\mathcal{A}_V(n,1/4,1/6)$. $\mathcal{A}_L$ makes $\mathcal{O}(n^4 \log(n))$ queries and $\mathcal{A}_V$ makes $\mathcal{O}(n^2 \log(n))$ queries. When acting on a unitary from $\mathcal{C}_n$, $\mathcal{A}_L$ produces a channel $\mathcal{E}$ within $\frac{1}{48n}$ diamond distance of the unitary with probability at least $5/6$. Using Lemma~\ref{lem:dist-avg-diam}, we see that the learned channel satisfies $d_{\mathrm{avg}}(\mathcal{E}, U(.)U^{\dag}) \leq d_{\diamond}(\mathcal{E}, U(.)U^{\dag}) \leq \frac{1}{48n}$. Conditioned on correctly learning, and due to the fact that the queried channel is a unitary, an application of $\mathcal{A}_V$ outputs \textbf{PASS} with probability at least $5/6$. Overall, a union bound tells us that $\mathcal{A}$ outputs \textbf{PASS} with probability at least $2/3$. 
    
    On the other hand, Theorem~\ref{thm:haar-bb-avg-lb}, with $\epsilon = 1/4$ and $\delta = 1/3$, tells us that when acting on a Haar-random unitary, the channel $\mathcal{E}^\prime$ produced by $\mathcal{A}_L$, which queries the unitary polynomially many times with just pure states, satisfies $d_{\mathrm{avg}}(\mathcal{E}^\prime, U(.)U^{\dag}) \leq 1/4$ with probability less than $1/3$.  Conditioned on $\mathcal{A}_L$ not learning the unknown unitary, the probability that an application of $\mathcal{A}_V$ outputs \textbf{PASS} is at most $1/6$. Thus, the probability that $\mathcal{A}$ outputs \textbf{PASS} over the Haar-measure is at most $1/3+1/6 = 1/2$.
    
    Thus, the advantage of $\mathcal{A}$ in distinguishing $\mathcal{C}_n$ from $\mu_H$ is at least $2/3-1/2 = 1/6$, which is non-negligible. Moreover, as both $\mathcal{A}_L$ and $\mathcal{A}_V$ run in $\mathsf{poly}(n)$ time, $\mathcal{A}$ is a computationally efficient distinguisher.
\end{proof}

\section{Outlook}
\label{sec:conclusion}
We have used the natural noise-tolerance of the QPSQ model to demonstrate the significance of statistical data for studying the learnability of quantum circuits in the near term.  Our method for characterizing the noise from the entangled statistics applies to global depolarizing noise, and opens up an important line of research - \textit{What kinds of noise be efficiently benchmarked using just statistical queries?}.

Together, the natural noise tolerance of our access model and the benchmarking method provide a useful framework for developing robust algorithms. By developing algorithms that only require statistical data, one can efficiently make them robust. We have shown how this can be done with the learning algorithm for shallow quantum circuits from~\cite{huang2024learning}, by adapting it to our statistical query setting with only a linear overhead in query complexity. Developing other learning algorithms in this access model is a promising line of research towards robust learning.

In the statistical query setting, we have shown an \emph{average-case} query-complexity lower bound for random quantum circuits of logarithmic to linear depth. Our bound does not rule out the possibility that at logarithmic depth, one might be able to develop efficient learning algorithms that succeed \textit{on average}. Up to linear depth, our lower bound shows an exponential scaling of the query complexity with the depth. At greater depths, random quantum circuits converge to approximate $2$-designs. For such circuits, exponential hardness for statistical query learning has already been shown in~\cite{wadhwa2024learning}. Our learning algorithm and lower bound thus provide strong insights into the learnability of random quantum circuits from statistical queries across all depth regimes. 

While we have defined a new quantum statistical query oracle for learning unknown observables, we have only instantiated it abstractly, by simulating it as a part of our learning algorithm for quantum circuits. We believe this oracle can have much wider applicability, especially for learning from physical experiments. Benchmarking the behaviour of unknown physical apparatus from statistical data is a critical problem in quantum information, and we believe this oracle can prove quite useful in the theoretical study of such problems. The same goes for our multi-copy oracle for learning processes. By allowing multi-copy queries, we are able to model an access model with more generality, and it would be quite interesting to observe \emph{new separations between multi-copy and single-copy statistical query oracles}, similar to the separation for purity testing shown in~\cite{nietner2023unifying}.

Our lower bounds for shallow random quantum circuits only hold when the depth is at least logarithmic. As we do not provide a lower bound at constant depths, it is natural to wonder \textit{whether the $\mathsf{QPStat}$ algorithm of Theorem~\ref{thm:qpsq-shallow-learner} is optimal.} As our lower bound technique requires some level of indistinguishability between the outputs of $\mathsf{QPStat}$ queries, we believe constant-depth circuits may not be sufficiently scrambled for this technique to provide meaningful lower bounds, and it might be necessary to develop novel techniques to obtain such a result.

Finally, we have shown an important limitation in constructing pseudorandom unitaries. Our result on the depth requirement for PRUs is not surprising. Most candidate constructions~\cite{metger2024simple,chen2024efficient} consist of circuits with a polynomial depth. While a lot of work is actively being done on proving the security of these constructions, proving stronger bounds on the depth necessary to achieve pseudorandomness would be a crucial result. Our method of combining the building blocks of learning and verification algorithms indicates a potential technique for achieving such results. 

\subsection*{Acknowledgements}
The authors acknowledge the support of the Quantum Advantage Pathfinder (QAP), with grant reference EP/X026167/1 and the UK Engineering and Physical Sciences Research Council. The authors thank anonymous reviewers of the conferences YQIS and QTML for helpful comments.

\bibliographystyle{unsrt}
\bibliography{biblio}

\appendix
\section{Estimating the noise with a single query}
\label{app:noise-estimation}
\begin{proof}[Proof of Theorem~\ref{thm:qpsq-noise-estim}]
Recall from the proof sketch that we make the following query,
\begin{equation}
    \alpha \leftarrow \mathsf{2QPStat}_{\Lambda(\mathcal{U})}(|0\rangle\langle0|^{\otimes 2}, \mathbb{F}, \tau),
\end{equation}
and the purity of the output state is
\begin{equation}
    \Tr((\rho^{out})^2) = 1 - (2\gamma-\gamma^2)(1-\frac{1}{2^n}).
\end{equation}
Denote by $f : [0,1] \rightarrow [0,1], f(\gamma) = 2\gamma-\gamma^2$. Note this function is strictly increasing for $\gamma \in [0,1)$. We have $f^{-1}(y) = 1 - \sqrt{1-y}$. Then, we have 
\begin{equation}
    \alpha \in \left[ 1-f(\gamma)(1-2^{-n})-\tau, 1-f(\gamma)(1-2^{-n})+\tau\right].
\end{equation}
Equivalently,
\begin{equation}
    f(\gamma) \in \left[ 
        \frac{1-\alpha-\tau}{1-2^{-n}}, \frac{1-\alpha+\tau}{1-2^{-n}}
    \right].
\end{equation}
Therefore, we obtain a range for the depolarizing strength $\gamma$
\begin{equation}
\label{eq:gamma-bound}
    \gamma \in \left[ 
        f^{-1}\left(\frac{1-\alpha-\tau}{1-2^{-n}}\right),  f^{-1}\left(\frac{1-\alpha+\tau}{1-2^{-n}}\right)
    \right].
\end{equation}
While we have an estimate of the noise strength, we need to estimate the diamond distance between the noiseless and noisy channels.

\begin{align}
    \norm{\mathcal{U}-\Lambda(\mathcal{U})}_\diamond &= \norm{\mathcal{U}-\Lambda(\gamma) \circ \mathcal{U}}_\diamond \\
    &= \norm{\mathcal{I}-\Lambda(\gamma)}_\diamond \\
    &= \underset{\rho}{\mathrm{max}} \norm{\mathcal{I} \otimes \mathcal{I}(\rho) - \Lambda(\gamma) \otimes \mathcal{I}(\rho) }_1 \\ &= \underset{\rho}{\mathrm{max}} \norm{\rho - (1-\gamma)\rho - \gamma(\Phi \otimes \mathcal{I})(\rho)}_1 \\ &= \gamma \underset{\rho}{\text{ }\mathrm{max}} \norm{\rho - \Phi \otimes \mathcal{I}(\rho)}_1 
    \\ &=\gamma \norm{\mathcal{I}-\Phi}_\diamond, \label{eq:noise-noiseless-half-bound}
\end{align}
where the second equality follows from the unitary invariance of the diamond norm and $\Phi$ is the maximally depolarizing channel.
From Lemma~\ref{lem:diamond-distance-dep}, we know that for any unitary channel $\mathcal{U}$,
\begin{equation}
\label{eq:unitary-far-dep}
    \norm{\mathcal{U}-\Phi}_\diamond \geq 2-\frac{2}{2^n}.
\end{equation}
As this lower bound is close to the maximum possible for the diamond norm, we use the upper bound
\begin{equation}
\label{eq:id-dep-diamond}
    \norm{\mathcal{I}-\Phi}_\diamond \leq 2.
\end{equation}
Combining (\ref{eq:gamma-bound}), (\ref{eq:noise-noiseless-half-bound}), (\ref{eq:unitary-far-dep}) and (\ref{eq:id-dep-diamond}), we obtain
\begin{equation}
    \label{eq:noise-noiseless-complete-bound}
    \norm{\mathcal{U}-\Lambda(\mathcal{U})}_\diamond \in 
    \left[
    2(1-2^{-n})f^{-1}\left(\frac{1-\alpha-\tau}{1-2^{-n}}\right),  2f^{-1}\left(\frac{1-\alpha+\tau}{1-2^{-n}}\right)
    \right].
\end{equation}
Denote
\begin{equation}
    [l,u] = \left[
    2(1-2^{-n})f^{-1}\left(\frac{1-\alpha-\tau}{1-2^{-n}}\right),  2f^{-1}\left(\frac{1-\alpha+\tau}{1-2^{-n}}\right)
    \right].
\end{equation}
Now, to show that this is a good estimate for the diamond norm between the noisy and noiseless unitaries, we show the choice of $\tau$ for which the difference between these bounds is small, i.e.
\begin{equation}
    u - l \leq \epsilon.
\end{equation}
First, we rewrite the above quantity as follows.
\begin{align}
        u - l &= 2\left(
        f^{-1}\left(\frac{1-\alpha+\tau}{1-2^{-n}}\right) - f^{-1}\left(\frac{1-\alpha-\tau}{1-2^{-n}}\right)
        \right) + \frac{2}{2^n}f^{-1}\left(\frac{1-\alpha-\tau}{1-2^{-n}}\right)
        \\&\leq 2\left(
        \frac{\sqrt{\alpha-2^{-n}+\tau}-\sqrt{\alpha-2^{-n}-\tau}}{\sqrt{1-2^{-n}}}
        \right) + \frac{2}{2^n}.
    \end{align}
Now, we will focus on the numerator of the first term.
\begin{align}
        \sqrt{\alpha-2^{-n}+\tau}-\sqrt{\alpha-2^{-n}-\tau} &= \frac{(\alpha-2^{-n}+\tau)-(\alpha-2^{-n}-\tau)}{\sqrt{\alpha-2^{-n}+\tau}+\sqrt{\alpha-2^{-n}-\tau}}
        \\ &\leq \frac{2\tau}{2\sqrt{\alpha-2^{-n}-\tau}}
        \\& \leq \frac{\tau}{\sqrt{\Tr(\rho)^2-2^{-n}-2\tau}}
        \\& = \frac{\tau}{\sqrt{(1-\gamma)^2(1-2^{-n})-2\tau}},
    \end{align}
where the second inequality uses $\alpha \in [\Tr(\rho^2)-\tau, \Tr(\rho^2)+\tau]$.
Suppose we start with some initial, loose upper bound on $\gamma \leq \gamma_u$. For instance, $\gamma_u = 0.5$.
Then, choose
\begin{equation}
     \tau = \frac{(1-\gamma_u)\epsilon}{4}.
\end{equation}
Denote by $C$
\begin{equation}
    C = \frac{\epsilon}{2(1-\gamma_u)}.
\end{equation} 
For $\epsilon \leq (1-\gamma_u)$, we have $C \leq 0.5$. Further,
\begin{equation}
    \tau = \frac{C(1-\gamma_u)^2}{2} \leq \frac{C(1-\gamma)^2}{2}.
\end{equation}
Now, for this value of $\tau$, 
\begin{align}
        \sqrt{\alpha-2^{-n}+\tau}-\sqrt{\alpha-2^{-n}-\tau} & \leq \frac{C(1-\gamma_u)^2}{2(1-\gamma)\sqrt{1-2^{-n}-C}} \\& \leq C(1-\gamma_u),
     \end{align}
where we use $1-\gamma_u \leq 1-\gamma$ and $\sqrt{1-2^{-n}-C} \geq 1/2, \forall n \geq 2, C \leq 1/2$ in the last inequality. Thus,
\begin{align}
        u -l  & \leq 2\frac{C(1-\gamma_u)}{\sqrt{1-2^{-n}}} - \frac{2}{2^n} \\&= \frac{\epsilon}{\sqrt{1-2^{-n}}}  - \frac{2}{2^n}
        \\&= \frac{\epsilon(\sqrt{1+2^{-n}})}{1-2^{-n}} - \frac{2}{2^n}
        \\&\leq \frac{\epsilon(1+2^{-n-1})}{1-2^{-n}} - \frac{2}{2^n}
        \\&= \frac{\epsilon(2^n+1/2)-2(1-2^{-n})}{2^n-1}
        \\&\leq \epsilon,
    \end{align}
where the second to last line uses $\sqrt{1+x} \leq 1+x/2 \text{ } \forall x \geq 0$ and the last inequality holds for all $n\geq 2, \epsilon \leq 1$.
\end{proof}

\section{Bounding variance of brickwork random circuits}
\label{app:shallow-var}
We will bound the variance of $\Tr(OU\rho U^\dag)$ over brickwork random quantum circuits using arguments from prior work~\cite{nietner2023average, hunter2019unitary,barak2020spoofing}. Their arguments involve representing the second-order moment in the form of a tensor network diagram, mapping the diagram to a statistical mechanics model, and then counting domain walls over this model. We will use the following lemma adapted from the arguments of~\cite{barak2020spoofing}, and refer to~\cite{hunter2019unitary, barak2020spoofing} for the diagrams and the detailed mapping. 
\begin{lemma}[Adapted from~\cite{barak2020spoofing}]
\label{lem:brqc-moment-haar}
        The second-order moments of brickwork random circuits can be written as
        \begin{equation}
            \underset{U \sim \mathsf{RQC}(n,d)}{\mathbf{E}} \left[\Tr(OU\rho U^\dag)^2\right] = f_1(\rho) f_2(n,d) f_3(O),
        \end{equation}
        for some functions $f_1,f_3 : \mathbb{C}^{2^n \times 2^n}\rightarrow \mathbb{R}$, and $f_2 : \mathbb{Z} \times \mathbb{Z \rightarrow \mathbb{R}}$, such that
        \begin{equation}
            f_2(n,d) \leq \left(1+\left(\frac{4}{5}\right)^d\right)^{n/2} \lim_{d^* \rightarrow \infty} f_2(n,d^*).
        \end{equation}
\end{lemma}
We can now prove Lemma~\ref{lem:shallow-variance}.
\begin{proof}[Proof of Lemma~\ref{lem:shallow-variance}]
    Recall that our goal is to show a bound on the variance of $\Tr(OU\rho U^\dag)$ where $U$ is a brickwork random quantum circuit. To this end, we bound the second moment using Lemma~\ref{lem:brqc-moment-haar}.
    \begin{align}
         \underset{U \sim \mathsf{RQC}(n,d)}{\mathbf{E}} \left[\Tr(OU\rho U^\dag)^2\right] &\leq  \left(1+\left(\frac{4}{5}\right)^d\right)^{n/2} f_1(\rho) \left( \lim_{d^* \rightarrow \infty}f_2(n,d^*)\right) f_3(O) \\& = \left(1+\left(\frac{4}{5}\right)^d\right)^{n/2} \lim_{d^* \rightarrow \infty} \underset{U \sim \mathsf{RQC}(n,d^*)}{\mathbf{E}} \left[\Tr(OU\rho U^\dag)^2\right]
         \\&= \left(1+\left(\frac{4}{5}\right)^d\right)^{n/2}  \underset{U \sim \mu_H}{\mathbf{E}} \left[\Tr(OU\rho U^\dag)^2\right],
    \end{align}
    where we use Lemma~\ref{lem:brqc-moment-haar}, and the fact that BRQCs converge to the Haar measure at infinite depth in the last step.
    Now, we use Lemma~\ref{lem:haar-moments} and the fact that BRQCs form exact 1-designs to compute the variance.
    \begin{align}
           \underset{U \sim \mathsf{RQC}(n,d)}{\mathbf{Var}}\left[\Tr(OU\rho U^\dag)\right] &= 
           \underset{U \sim \mathsf{RQC}(n,d)}{\mathbf{E}} \left[\Tr(OU\rho U^\dag)^2\right] - \left(\underset{U \sim \mathsf{RQC}(n,d)}{\mathbf{E}} \left[\Tr(OU\rho U^\dag)\right]\right)^2 \\&\leq 
           \left(1+\left(\frac{4}{5}\right)^d\right)^{n/2}\underset{U \sim \mu_H}{\mathbf{E}} \left[\Tr(OU\rho U^\dag)^2\right] \\&-\left(\underset{U \sim \mu_H}{\mathbf{E}} \left[\Tr(OU\rho U^\dag)\right]\right)^2.
        \end{align}
    Note that
    \begin{equation}
        \left(1+\left(\frac{4}{5}\right)^d\right)^{n/2} \leq \exp\left(\frac{n\left(4/5\right)^d}{2}\right).
    \end{equation}
    For $d \geq \frac{\log n}{\log 5/4},$ we have $ \frac{n(4/5)^d}{2} \leq 1/2$. We then use the fact that $e^x \leq 1+2x \text{ }, \forall \text{ } 0 \leq x \leq 1 $ to show
    \begin{equation}
        \left(1+\left(\frac{4}{5}\right)^d\right)^{n/2} \leq \left(1+n\left(\frac{4}{5}\right)^d\right).
    \end{equation}
    Denote $N = 2^n$. Now, we can bound the variance as:
    \begin{align}
             \underset{U \sim \mathsf{RQC}(n,d)}{\mathbf{Var}}\left[\Tr(OU\rho U^\dag)\right] &\leq \left(1+n\left(\frac{4}{5}\right)^d\right)\underset{U \sim \mu_H}{\mathbf{E}} \left[\Tr(OU\rho U^\dag)^2\right] - \left(\underset{U \sim \mu_H}{\mathbf{E}} \left[\Tr(OU\rho U^\dag)\right]\right)^2
           \\&=  \underset{U \sim \mu_H}{\mathbf{Var}}\left[\Tr(OU\rho U^\dag)\right]+n\left(\frac{4}{5}\right)^d\underset{U \sim \mu_H}{\mathbf{E}} \left[\Tr(OU\rho U^\dag)^2\right]
           \\&\leq \frac{1}{N+1} + n\left(\frac{4}{5}\right)^d\left(\left(\frac{N-\Tr(\rho^2)}{N(N^2-1)}\right)\Tr(O)^2 + \left(\frac{N\Tr(\rho^2)-1}{N(N^2-1)}\right)\Tr(O^2)\right) 
           \\&\leq \frac{1}{N+1} + n\left(\frac{4}{5}\right)^d\left(\frac{\Tr(O)^2}{N^2} + \frac{\Tr(O^2)}{N(N+1)}\right)
           \\&\leq \frac{1}{N+1} + n\left(\frac{4}{5}\right)^d\left(1+\frac{1}{N+1} \right) 
           \\&= n\left(\frac{4}{5}\right)^d + \frac{1}{2^n+1}\left(1 + n\left(\frac{4}{5}\right)^d\right),
        \end{align}
    where the second inequality follows from Lemma~\ref{lem:haar-moments}, the third inequality follows from the fact that $1/N \leq \Tr(\rho^2) \leq 1$. To obtain the fourth inequality, observe that for $\norm{O}_\infty \leq 1, |\Tr(O)| \leq N,$ and $ \Tr(O^2) \leq N$. The final equation is obtained by rearranging the terms and substituting $N = 2^n$. Now, to obtain the desired bound on the variance, we need 
    \begin{equation}
        n\left(\frac{4}{5}\right)^d = \Omega(2^{-n})
    \end{equation}
    which is satisfied when
    \begin{equation}
        d \leq \frac{n+\log_2(n)}{\log_2(5/4)}
    \end{equation}
    Thus, the desired bound on the variance is obtained for the outlined depth range, concluding the proof.
\end{proof}

\section{Average-case hardness for Haar-random unitaries}
\label{sec:haardness}
Our proof will build upon the technique of~\cite{arapinis2021quantum}, where the authors proved a bound on the ability of an adversary to predict the output of a Haar-random unitary on a Haar-random state with high fidelity. We will also use the following lemma on the average infidelity of any CPTP map with Haar-random unitaries on average. 
\begin{lemma}[Quantum no-free lunch theorem with no samples]
\label{lem:qnflt-no-samples}
Any CPTP map $\mathcal{E} : \mathcal{S}_N \rightarrow \mathcal{S}_N$ has high $d_{\mathrm{avg}}$ from Haar-random unitaries on average.
\begin{equation}
    \underset{U \sim \mu_H}{\mathbf{E}}d_{avg}(\mathcal{E}, U(.)U^\dag) = 1-\frac{1}{N}
\end{equation}
\begin{proof}
    \begin{align}
            \underset{U \sim \mu_H}{\mathbf{E}}d_{\mathrm{avg}}(\mathcal{E}, U(.)U^\dag) & = 1 - \underset{U \sim \mu_H}{\mathbf{E}}\left[\underset{|\psi\rangle \sim \mu_S}{\mathbf{E}}F\left(\mathcal{E}(|\psi\rangle\langle\psi|), U|\psi\rangle\langle\psi|U^\dag\right)\right] \\&=
            1-\underset{|\psi\rangle \sim \mu_S}{\mathbf{E}}\left[\underset{U \sim \mu_H}{\mathbf{E}}F\left(\mathcal{E}(|\psi\rangle\langle\psi|), U|\psi\rangle\langle\psi|U^\dag\right)\right]  \\&=
            1 - \underset{|\psi\rangle \sim \mu_S}{\mathbf{E}}\left[\underset{U \sim \mu_H}{\mathbf{E}}\Tr(\mathcal{E}(|\psi\rangle\langle\psi|)U|\psi\rangle\langle\psi|U^\dag)\right] 
            \\&= 1 - \underset{|\psi\rangle \sim \mu_S}{\mathbf{E}}\left[\Tr\left(\mathcal{E}(|\psi\rangle\langle\psi|)\underset{U \sim \mu_H}{\mathbf{E}}[U|\psi\rangle\langle\psi|U^\dag]\right)\right]
            \\&= 1 - \underset{|\psi\rangle \sim \mu_S}{\mathbf{E}}\left[\Tr\left(\mathcal{E}(|\psi\rangle\langle\psi|)\frac{I}{N}\right)\right]
            \\& = 1 - \underset{|\psi\rangle \sim \mu_S}{\mathbf{E}}\left[\frac{1}{N}\right]
            \\& = 1-\frac{1}{N},
        \end{align}
    where the third equality uses the property of the fidelity when one of the states is pure, the fourth equality uses the linearity of expectation and trace, the fifth equality follows from Lemma~\ref{lem:haar-moments}, and the second to last equality follows from the fact that $\mathcal{E}$ is trace-preserving.
\end{proof}
\end{lemma}
The lemma can be interpreted as follows: any learning algorithm for a Haar-random unitary that makes \textit{no queries}, can only guess either a fixed channel or a channel from some fixed distribution. Lemma~\ref{lem:qnflt-no-samples} then tells us that such an algorithm will have high error on average over all Haar-random unitaries. This interpretation is similar to the original QNFLT from~\cite{poland2020no}. In fact, by setting the number of queries to 0 in the original theorem, we obtain the same bound as the original QNFLT. Now, we will prove Theorem~\ref{thm:haar-bb-avg-lb}.
\begin{proof}[Proof of Theorem~\ref{thm:haar-bb-avg-lb}]
    Suppose the learner makes queries $\{|\psi_{in}^i\rangle\}_{i = 1}^q$ and receives states $\{|\psi_{out}^i\rangle\}_{i = 1}^q$, with $|\psi_{out}^i\rangle = U|\psi_{in}\rangle$. We strengthen the access of the learner by assuming the learner also has access to the classical descriptions of the states. We will show the lower bound for this strengthened learner, which will then hold for the original setting as well.
    
    Given the classical description, the learner has full knowledge of the action of $U$ over the subspace spanned by the input states. Denote the Hilbert space over $n$-qubit states as $\mathcal{H}^N$, where $N = 2^n$. Denote the space of states with non-zero overlap with the input states as $\mathcal{H}^q$, and the space of states orthogonal to all input states as $\mathcal{H}^{q^\perp}$. Denote by $\mu_H^\prime$ ($\mu_S^\prime$) the Haar measure over unitaries (states) on the Hilbert space $\mathcal{H}^{q^\perp}$. Similar to~\cite{arapinis2021quantum}, we strengthen the learner beyond the assumption of~\cite{poland2020no}, by giving it perfect fidelity on every state in $\mathcal{H}^q$. Thus, as long as the overlap of a state with any $|\psi_{in}^i\rangle$ is non-zero, even if it's arbitrarily small, we assume the learner succeeds perfectly. Thus, the learner has no error on states in $\mathcal{H}^q$. On the other hand, for states in $\mathcal{H}^{q^\perp}$, the action of the unitary is completely random. Denote by $d_\mu$ the average infidelity between two channels over states sampled from some measure $\mu$. In particular, when $\mu = \mu_S$, $d_\mu$ is $d_{\mathrm{avg}}$. Now, for a learner in the original setting, with learned channel $\mathcal{E}$, we bound the expected error as 
    \begin{equation}
        \underset{U \sim \mu_H}{\mathbf{E}}[d_{\mathrm{avg}}(U, \mathcal{E})] \geq \underset{|\psi\rangle \sim \mu_S}{\mathbf{Pr}}\left[|\psi\rangle \notin \mathcal{H}^{q^\perp}\right] 0 + \underset{|\psi\rangle \sim \mu_S}{\mathbf{Pr}}\left[|\psi\rangle \in \mathcal{H}^{q^\perp}\right]
        \underset{U^\prime \sim \mu_H^\prime}{\mathbf{E}}\left[d_{\mu_S^\prime}(U^\prime, \mathcal{E}|_{\mathcal{H}^{q^\perp}})
        \right],
    \end{equation}
    where $\mathcal{E}|_{\mathcal{H}^{q^\perp}}$ denotes $\mathcal{E}$ restricted to input states from $\mathcal{H}^{q^\perp}$. The expected error over $\mu_H^\prime$ is now precisely given by Lemma~\ref{lem:qnflt-no-samples} for dimension $N-q$. Next, we compute the probabilities in the equation. As shown in~\cite{arapinis2021quantum}, the probability of $|\psi\rangle$ belonging to the subspace $\mathcal{H}^{q^\perp}$ is given by the ratio of the dimensions, i.e. 
    \begin{equation}
        \underset{|\psi\rangle \sim \mu_S}{\mathbf{Pr}}\left[|\psi\rangle \in \mathcal{H}^{q^\perp}\right] = \frac{N-q}{N}.
    \end{equation}
    Thus, we obtain
    \begin{equation}
        \label{eq:avg-risk-lower bound}
         \underset{U \sim \mu_H}{\mathbf{E}}[d_{\mathrm{avg}}(U, \mathcal{E})] \geq \frac{N-q}{N}\left(1-\frac{1}{N-q}\right) = 1 - \frac{q+1}{N}.
    \end{equation}
    Next, we will use a crude upper bound on the average error.
    \begin{align}
        \underset{U \sim \mu_H}{\mathbf{E}}[d_{\mathrm{avg}}(U, \mathcal{E})] &\leq \epsilon \cdot \underset{U \sim \mu_H}{\mathbf{Pr}}[d_{\mathrm{avg}}(U, \mathcal{E})\leq \epsilon] + 1 \cdot \underset{U \sim \mu_H}{\mathbf{Pr}}[d_{\mathrm{avg}}(U, \mathcal{E}) > \epsilon]
        \\&= 1 - (1-\epsilon)\left(\underset{U \sim \mu_H}{\mathbf{Pr}}[d_{\mathrm{avg}}(U, \mathcal{E})\leq \epsilon]\right),
    \end{align}
    where the first inequality uses the fact that the average error is upper bounded by 1. Now, for success probability at least $1-\delta$, we have the upper bound
    \begin{equation}
        \label{eq:avg-risk-upper-bound}
        \underset{U \sim \mu_H}{\mathbf{E}}[d_{\mathrm{avg}}(U, \mathcal{E})] \leq 1-(1-\epsilon)(1-\delta).
    \end{equation}
    Combining the upper bound (\ref{eq:avg-risk-upper-bound}) and lower bound (\ref{eq:avg-risk-lower bound}), we obtain
    \begin{equation}
        q \geq N(1-\epsilon)(1-\delta)-1,
    \end{equation}
    as desired.
\end{proof}

\end{document}